\documentclass{aa}  

\usepackage{graphicx}
\usepackage{txfonts}
\def\simless{\mathbin{\lower 3pt\hbox
     {$\rlap{\raise 5pt\hbox{$\char'074$}}\mathchar"7218$}}}   
\def\simmore{\mathbin{\lower 3pt\hbox
     {$\rlap{\raise 5pt\hbox{$\char'076$}}\mathchar"7218$}}}   

\def\msun{~{\rm M}_\odot}

\begin{document}

\title{Illumination of the accretion disk in black hole binaries: 
An extended jet as the primary source of hard X-rays}

\subtitle{}

\author{P. Reig\inst{1,2}
\and
N. D. Kylafis\inst{2,1}
}
\institute{Institute of Astrophysics, Foundation for Research and Technology-Hellas, 71110 Heraklion, Crete, Greece\\
\email{pau@physics.uoc.gr}
\and
University of Crete, Physics Department, 70013 Heraklion, Crete, Greece\\
\email{kylafis@physics.uoc.gr}           
}

\date{}

 
\abstract
{The models that seek to explain the reflection spectrum in black hole
binaries usually invoke a point-like primary source of hard X-rays. 
This source illuminates the accretion disk and gives rise to the discrete 
(lines) and continuum-reflected components.}
{The main goal of this work is to investigate whether the extended, mildly 
relativistic jet that is present in black hole binaries
in the hard and hard-intermediate states
is the hard X-ray source that illuminates the accretion disk.}
{We use a Monte Carlo code that simulates the process of inverse Compton 
scattering in a mildly relativistic jet rather than in 
a ``corona'' of some sort. 
Blackbody photons from the thin accretion disk are injected at the base 
of the jet and interact with the energetic electrons that move outward with 
a bulk velocity that is a significant fraction of the speed of light.} 
{Despite the fact that the jet moves away from the disk at a mildly relativistic
speed, we find that approximately $15-20$\% of the input soft photons
are scattered, after Comptonization, back toward the accretion disk. 
The vast majority of the Comptonized,
back-scattered photons escape very close to the 
black hole ($h\simless 6 r_g$, where $r_g$ is the gravitational radius),
but a non-negligible amount escape at a wide range of heights. At high heights,
$h\sim 500-2000\,r_g$, the distribution falls off rapidly. The high-height cutoff
strongly depends on the width of the jet at its base and is almost insensitive to 
the optical depth. The disk illumination spectrum
is  softer than the direct jet spectrum of the radiation that escapes
in directions that do not encounter the disk.
}
{We conclude that an extended jet is an excellent candidate source of
hard photons in reflection models.
}


\authorrunning{Reig \& Kylafis 2020}

\titlerunning{Disk irradiation in BHBs }

   \maketitle
%

\begin{figure*}
\centering
\includegraphics[width=14cm]{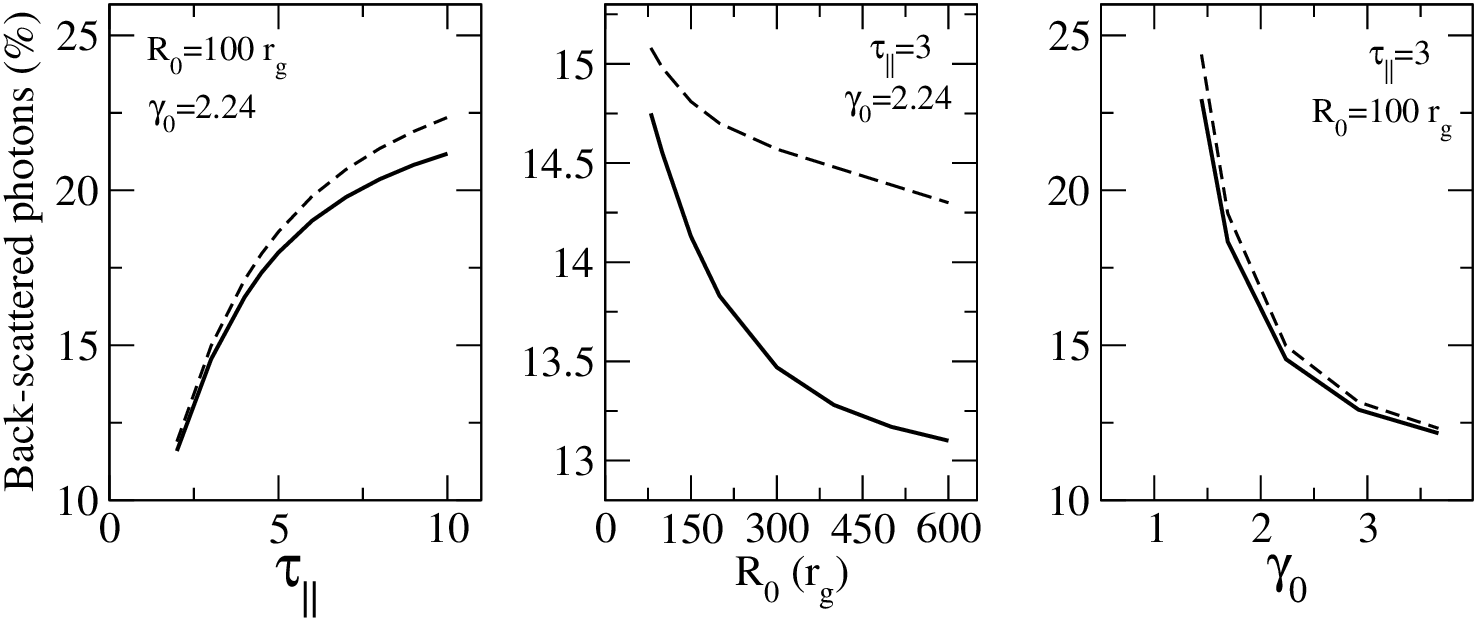}
\caption{Fraction of input photons that are back-scattered (dashed lines) and
also illuminate the accretion disk (solid lines) as a function of optical 
depth (left panel), jet width (middle panel), 
and the Lorentz factor (right panel).
}
\label{downphot}
\end{figure*}

\section{Introduction}

The observed X-ray spectral continuum (0.1--200 keV) of black hole binaries
(BHBs) exhibits a power law with a high-energy exponential cutoff. This
power law is affected by interstellar photoelectric absorption at low  energies
(typically below 2 keV).  The exponential 
cutoff  varies in a broad range of energies, from a
few tens to a few hundred keV, and the photon-number power-law index $\Gamma$
varies from 1.2 to 3. The values of these parameters define spectral states that are broadly
known as hard ($\Gamma < 2$) and soft ($\Gamma > 2.5$). Other parameters related
to the time variability of the source, such as the shape and strength of the
power spectrum and the frequency of quasi-periodic oscillations,  contribute to
the characterization of these spectral states \citep{mclintock06,belloni10}.
Thermal emission from the accretion disk adds up to the observed spectrum.  
Superimposed on the continuum, discrete components are present, and they are
attributed to reflection in the accretion disk \citep[see,
e.g.,][]{fabian10,bambi21}. The
most prominent feature of the reflected spectrum is the iron emission line at
6.4 keV. Therefore, the observer sees a combination of direct hard emission and
reflected emission.

Although there is a general consensus that the hard X-rays result from the inverse
Compton  scattering of low-energy photons from the accretion disk by high-energy
electrons, the physical nature and the geometry of the primary source of the accretion disk irradiation is still a matter of debate.  Because of the
complexity of the physics involved, many authors consider  a simple geometry in
which the illuminating continuum is assumed to be emitted isotropically from a
point  source on the rotational axis at height $h$ above the black hole. This
model is known as the lamp post model, and it is often used to model the
reflection features and determine the black hole  spin \citep[see,
e.g.,][]{reynolds14} in accreting black holes present both in active galactic
nuclei  \citep{martocchia96,martocchia02,miniutti03,emmanoulopoulos14,zoghbi20}
and X-ray binaries \citep{duro16,vincent16,basak16,garcia19}.

The purpose of this work is to investigate  whether the jet, moving away  from
the black hole at  a mildly relativistic velocity in the hard and
hard-intermediate states, can be the source of the hard  photons that illuminate
the accretion disk.  Here we are interested in the irradiation of the disk
before it is reflected. Computing the reflected spectrum is beyond of the scope
of this work.

\section{The jet model}

Hot-inner-flow Comptonization models have been invoked to explain the  hard
X-ray emission of BHBs, and they have been very successful at reproducing the
observations \citep[e.g.,][]{esin97,done07}. Propagating-fluctuation models have
been used to explain the time lag of hard X-rays with respect to softer ones
\citep{lyubarskii97,kotov01,arevalo06,rapisarda17}, and they have also been
quite successful.   However, these models are independent of one another and
have not been able to explain the observed correlation between the spectrum and
the time lags \citep{vignarca03,pottschmidt03,altamirano15,reig18}. For this
reason, we favor Comptonization in the jet, which can explain not only the
spectra, but also the time lags and the correlation  between the two
\citep{reig15,reig18,kylafis18}. It is natural to expect a correlation between
the spectrum and the time lags when both are produced by the same mechanism,
namely Comptonization. In addition, we remark that Comptonization in the jet
quantitatively explains the type-B quasi-periodic oscillations (QPO) that have been
observed in GX 339-4 \citep{kylafis20}.

Our jet model was described in \citet[][see also \citealt{kylafis08}]{reig19},
where we also gave a justification of the parameters used. Here, we only provide
the essential points.

We assume a parabolic jet of radius $R(z) = R_0 (z/z_0)^{1/2}$,
where $R_0$ is the radius at the base of the jet,  
with an acceleration zone at its bottom, from height $z_0=5r_g$ 
to height $z_1=50r_g$, and a constant speed of $v_0=0.8c$ above this. 
Here, $r_g=GM/c^2$ is the gravitational radius and the height of the jet is 
$H= 10^5 r_g$. The flow speed in the acceleration zone is modeled by 
$v_{\parallel}(z) =(z/z_1)^p ~ v_0$, where $p=1/2$.

The Thomson optical depth $\tau_{\parallel}$ along the axis of the jet  is
\begin{equation}
\tau_{\parallel}=\int_{z_0}^{H}{\sigma_{\rm T} n_e(z) dz}.
\end{equation}

At any height $z$, the Thomson optical depth above this height is 
\begin{equation}
\tau_{\rm out}(z)=\int_{z}^{H}{\sigma_{\rm T} n_e(z) dz},
\end{equation}
while the Thomson optical depth perpendicular to the jet axis 
at height $z$ is 
\begin{equation}
\tau_{\perp}(z)=\sigma_{\rm T} n_e(z) R_0 (z/z_0)^{1/2}.
\end{equation}

The electrons are assumed to move on helical orbits around the magnetic field,
with velocity components $v_\parallel(z)$ and $v_\perp=0.4c$. 
Their Lorentz factor is $\gamma(z)=1/\sqrt{1-(v_\parallel^2+v_\perp^2)/c^2}$,
and in the coasting region of the flow, $z>z_1$, it is 
$\gamma_0=1/\sqrt{1-(v_0^2+v_\perp^2)/c^2}$.

We assume a 10$\msun$ non-spinning black hole. The input source of photons has a
blackbody distribution with $kT=0.2$ keV, presumably coming from the inner part
of the accretion disk. The disk is truncated well beyond the innermost stable
circular orbit (ISCO). We note that our results do not depend on the truncation
radius of the disk as long as enough photons enter the jet and undergo
Comptonization.

\section{Results}

Our Monte Carlo code records the number of photons that escape from the jet  as
a function of time (travel time in the Comptonizing medium), energy, and
direction (angle $\theta$ with respect to the jet axis). It also records the
height $h$ in the jet from which the photons escape.  Thus, we are able to
compute not only the fractional number of photons escaping directly toward the
observer and those that are back-scattered and hit the accretion disk, but also
their  spectrum and the height distribution of the points of escape.

The input parameters in our model calculations are $\tau_\parallel$, $R_0$, and
$\gamma_0$. The parameter space that we consider here for the optical depth and
jet width is determined from the range of values of the parameters that produce
good fits to the time lag--photon index correlation in GX\,339--4
\citep{kylafis08}, namely $2 \leq \tau_\parallel \leq 5$ and $100 \, r_g \leq
R_0 \leq 500 \,r_g$. For $\gamma_0$, we consider the results from
\citet{saikia19}, who used the near infrared excess observed in BHBs to
constrain the Lorentz factor in the jet in the range 1.3--3.5.

\subsection{Illumination of the accretion disk}

Figure~\ref{downphot} shows the fraction of back-scattered photons as a function
of optical depth (left panel), jet width (middle panel),  and Lorentz factor
(right panel).  In each case, one parameter is varied and the other two are kept
at reasonable values.  The solid lines give the fraction of back-scattered
photons that hit the disk between $R=R_{\rm ISCO}$ and $R=10^3 r_g$, while the
dashed lines give the fraction of back-scattered photons (i.e., with $\cos\theta
<0$).

It is evident from the left panel of Fig.~\ref{downphot} that,  for reasonable
values of the optical depth $\tau_\parallel$, the large majority of photons
escape in such directions that they will not  encounter the disk, which is expected.  What
is interesting is that a significant fraction of the escaping photons go back to
the disk,  despite the bulk motion of the electrons in the jet that  tends to
push them in the forward direction. The fraction of back-scattered photons
increases with optical depth, and the  relatively low bulk velocity in the
acceleration zone helps. The fraction of  back-scattered photons (dashed line)
is slightly larger than that of those that  hit the disk (solid line), and this is
expected too.

In the middle panel of Fig.~\ref{downphot}, we show the dependence  of the
back-scattered photons  on the size of the jet, as measured by  the radius $R_0$
at its base. An increase in $R_0$ results in a decrease in the back-scattered
fraction, albeit a small one.   For an increase by a factor of six in $R_0$, the
back-scattered fraction decreases by $\sim 5\%$ (dashed line), while the fraction
that hits the disk decreases by $\sim 10\%$ (solid line).  

In the right panel of Fig.~\ref{downphot}, one can see the dependence of the
back-scattered photons on the Lorentz factor $\gamma_0$ of the electrons in the
coasting region of the flow ($z>z_1$).  The variation of $\gamma_0$ is due to
the variation of the flow speed $v_0$, and $v_\perp$ is kept constant. As
expected, the fraction of back-scattered photons decreases rapidly with
increasing $\gamma_0$.  The leveling off at large $\gamma_0$ may seem
unphysical, but it is due to the scatterings that occur in the acceleration
zone. Irrespective of the final $v_0$, the parallel velocity is significantly
smaller than $v_0$ in the lower parts of the acceleration zone, and hence there are
no large variations of the Lorentz factor for the
cases considered in Fig.~\ref{downphot} in this region.

\begin{figure}
\centering
\includegraphics[width=8cm]{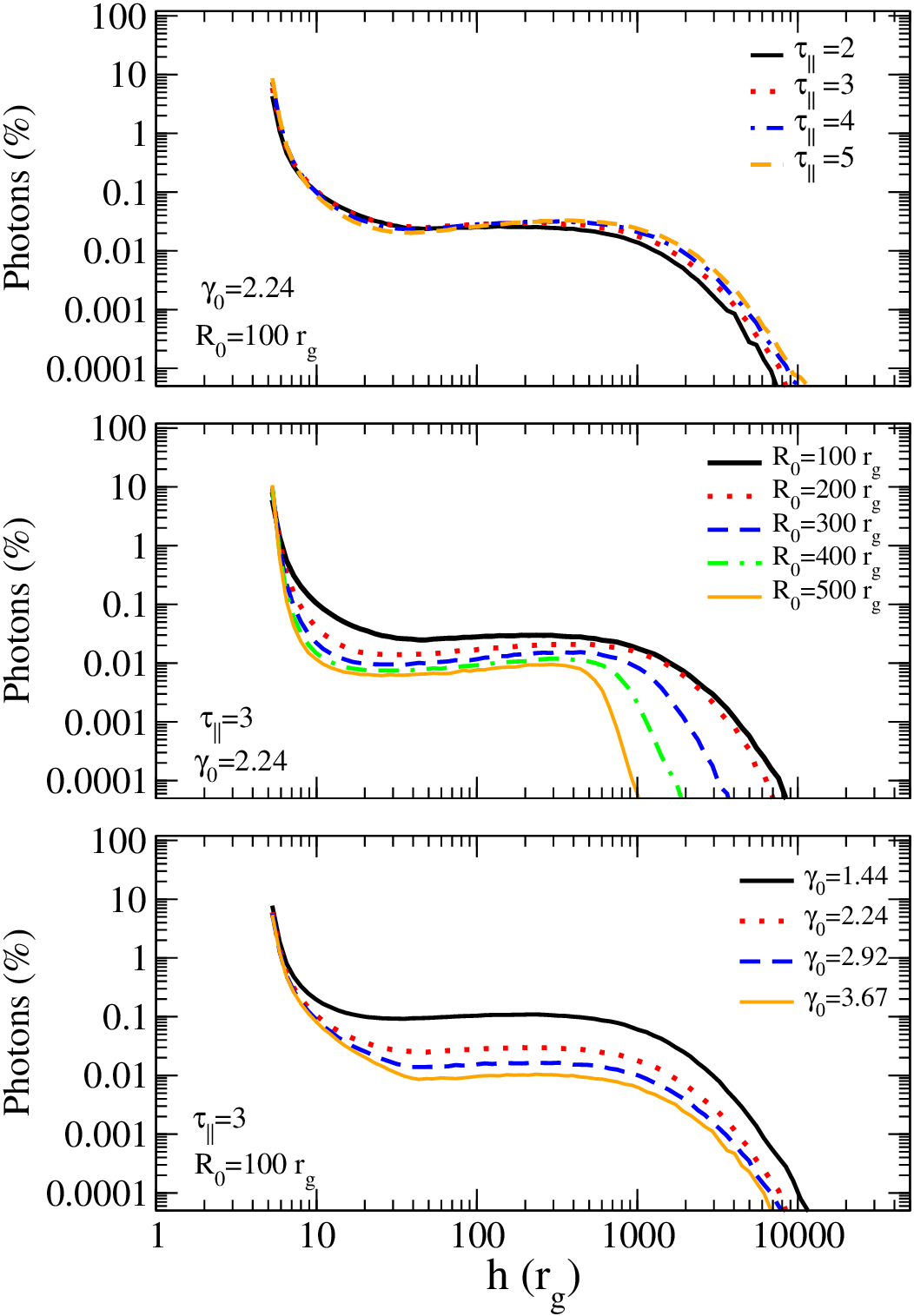}
\caption{Fraction of input photons that illuminate the disk 
 as a function of the height $h$ from where they escape the jet.
}
\label{height-dist-hitdisk}
\end{figure}

\subsection{Height distribution of back-scattered photons}
\label{hdist}

We now examine  the percentage of photons that have back-scattered and hit the
accretion disk  as a function of the height $h$ of the point of  last
scattering. The height distribution (Fig.~\ref{height-dist-hitdisk}) is
characterized by a sharp peak at a low height (within a few gravitational
radii), an extended plateau, and a high-height cutoff. In our jet model,
Comptonization can occur anywhere in the jet (not only at its base). Thus, it
is not surprising to see photons escaping from a large range of heights. 

Figure~\ref{height-dist-hitdisk} shows the results of running different models
with different optical depth $\tau_{\parallel}$, jet width $R_0$, and Lorentz
factor $\gamma_0$. In each model, we vary one parameter and keep the other two
at reasonable values. The results can be summarized as follows: {\em i)} the height
distribution does not depend strongly on optical depth (top panel), {\em ii)}
the cutoff at high $h$ moves to smaller values as $R_0$ increases (middle panel),
and {\em iii)} as $\gamma_0$ increases, the fraction of 
the photons back-scattered toward the disk at every $h$ decreases (bottom panel).

   \begin{figure}
   \centering
   \includegraphics[width=8cm]{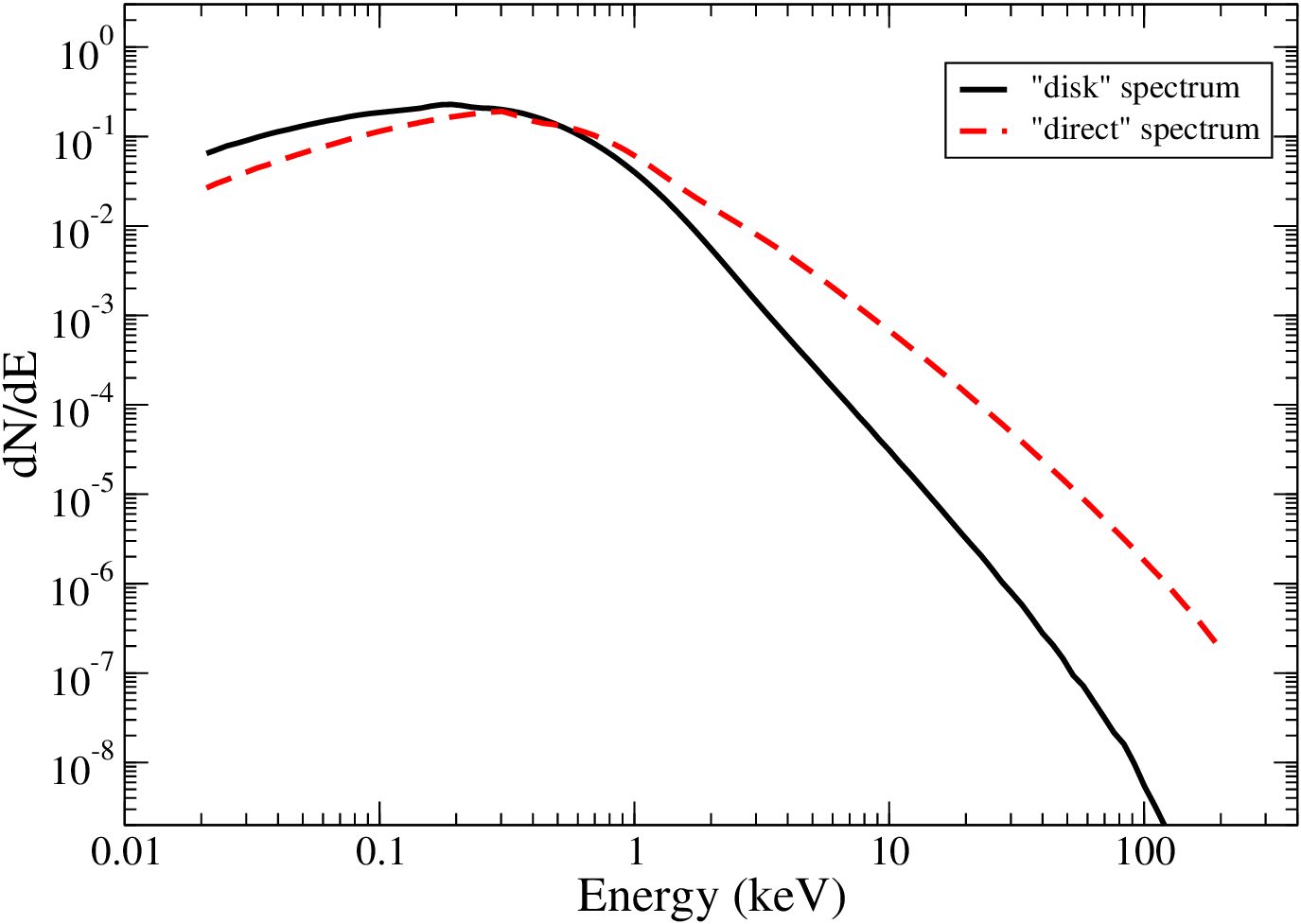}
      \caption{Comparison of the "direct" and "disk" spectra. The direct
    spectrum is the spectrum seen by observers at infinity at an inclination $\theta$,
    and the "disk" spectrum is the spectrum of the photons that are 
    back-scattered and illuminate the disk. The direct spectrum was computed for
    $0.7 < \cos \theta \le 0.8$). The parameters of the model used in this figure are: 
    $\tau_{\parallel}=3$, $R_0=80\,r_g$, and $\gamma_0=2.24$.}
         \label{disk-jet}
   \end{figure}

\subsection{Spectrum of the back-scattered photons}

Having demonstrated that a significant fraction of Comptonized photons move back
toward the disk,  we computed their spectrum.  Figure~\ref{disk-jet} compares 
the spectrum of the photons that illuminate the disk, the "disk" spectrum (solid
line), with the "direct" spectrum of the photons seen by observers  at angles
$\theta$, with respect to the jet axis, in the range $0.7 < \cos\theta \le 0.8$
(dashed line) or, equivalently, for a source inclination of $\sim 40^{\circ}$.
Both spectra can be fitted with power laws with a high-energy cutoff. The disk
spectrum is clearly softer than the direct spectrum. We emphasize that the disk
spectrum is not the spectrum of the reflected radiation.  It simply
represents the energy distribution of the photons that illuminate the disk. 

From the disk and the direct spectra, 
we  can compute the ratio of the illuminating and the 
directly observed radiation. This ratio is generally known as the reflection
fraction $R_F$, although different authors give different definitions depending
on whether the reflected flux, as opposed to the illuminating flux, is considered or
whether relativistic (i.e., light-bending) effects are taken into account 
\citep[see the appendix of ][for a detailed account of the reflection
fraction]{basak16}. In the lamp post
model, $R_F$ has been used to constrain the spin of the black hole
\citep{dauser14} and the geometry of the source of hard X-rays \citep{dauser16}.

Here we define $R_F$ simply as the ratio of the intensity  that illuminates the
disk to the intensity that directly reaches observers at infinity in a range of
observing angles $\theta$.   The reflection fraction is  then $R_F=I_{\rm
disk}/I_{\rm direct}$, where the intensities $I_{\rm disk}$ and $I_{\rm direct}$
are computed as 

\begin{equation}
I_i=\int_{E_{\rm min}}^{E_{\rm max}}{E \times (dN/dE)_i \times dE},
\end{equation}

\noindent where $dN/dE$ is the number of photons per unit energy in the spectrum
and the subscript $i$ refers to either "disk" or "direct."  We calculated
the reflection fraction in the energy range 1--100 keV, which is the range where
the power law dominates. We divided the disk into 13 radial zones and computed
the intensity of the photons that hit the disk between $R_i$ and $R_{\rm max}$,
where $R_i$ is the inner radius of each zone. In other words, we mimicked the
situation of a truncated disk whose inner radius decreases (i.e., the disk
is approaching the black hole). We note that because  the direct spectrum $I_{\rm
direct}$ depends on the angle $\theta$ between the jet axis and the observer
\citep{reig19,kylafis20}, so does the reflection fraction.

In the computation of the reflection fraction, we ignored relativistic
light-bending effects. In our model, we assume that photons are injected at the
sides of the jet at its base. Because $R_0$ is on the order of tens of
gravitational radii, the number of photons that pass near the black hole is
very small. Typically, the fraction of the input photons that enter the
sphere $r < 10\, r_g$  centered around the black hole is $\simless 0.01$\%
for $R_0 \simmore 50\, r_g$.

\begin{figure}
   \centering
   \includegraphics[width=8cm]{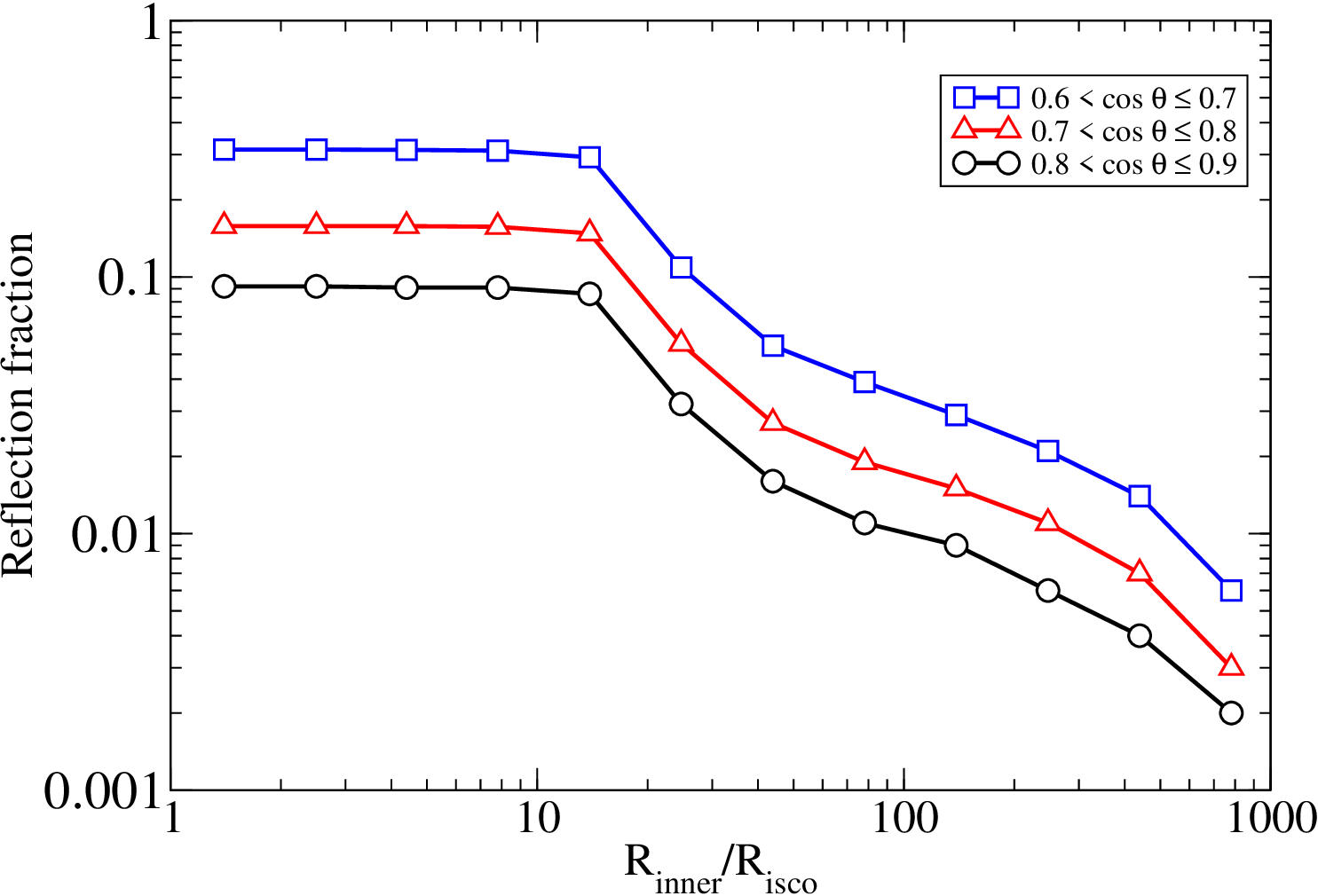}
      \caption{ Reflection fraction as a function of the inner radius
      for model parameters $\tau_{\parallel}=3$, $R_0=80\,
      r_g$, and $\gamma_0=2.24$.}
         \label{ref-frac-rad}
   \end{figure}

In Fig.~\ref{ref-frac-rad}, we show $R_F$ as a function of a disk inner radius for
three different inclinations,  $0.6 < \cos\theta \le 0.7$ (i.e., $\theta \sim
50^{\circ}$), $0.7 < \cos\theta \le 0.8$ (i.e., $\theta \sim 40^{\circ}$), and
$0.8 < \cos\theta \le 0.9$ (i.e., $\theta \sim 30^{\circ}$). The $R_F$ increases as
the inner disk radius $R_i$ decreases, and it increases as the inclination of the
system increases. The flat part of the curve reflects the fact that the inner
parts of the disk are scarcely illuminated by back-scattered photons.

\section{Discussion}

The reflection spectrum results from the reprocessing of hard X-ray photons by
the optically thick accretion disk. Therefore, we expect the properties of
the illuminating radiation to have a profound impact on the spectral features of
the reflection spectrum \citep{dauser13,garcia15a,steiner17}.  Because of the
complexity of the physics involved,  many authors assume a static  point
source that emits isotropically, as in the lamp post model or coronal
models \citep[see, e.g.,][]{vincent16}. In reality, the source is expected to be
variable, extended, anisotropic, and partly off-axis \citep{dauser13}.
Constraining the nature and geometry of the source of the hard X-rays that
illuminates the disk will allow us to consider more realistic models. The aim of this
work has been to investigate whether the jet itself can be the disk-illuminating
source.

In this section, we discuss the process of disk illumination as well  as the
results obtained in the previous section. We have shown that a significant
fraction of the photons return toward the accretion disk after being scattered
in the jet. We have also determined the height at which the photons that  are
back-scattered toward the disk escape from the jet. Finally, we have computed
the spectrum that  illuminates the disk. Below, we elaborate on all three of
these findings.

We have shown that, regardless of the optical depth, jet width, and jet velocity,
a significant fraction of the Comptonized photons are back-scattered and hit
the disk (Fig.~\ref{downphot}). The decrease in the number of back-scattered
photons as the optical depth decreases is expected because, for low optical
depths, the photons find it easier to move along the jet and travel longer
distances.  The higher up in the jet they travel, the less likely it is for the 
back-scattered photons to hit the disk. Likewise, a larger outflow velocity of the
electrons imposes a stronger forward motion on the photons and makes
back-scattering more difficult.  The decrease in the number of back-scattered
photons  as the jet width increases reflects the fact that photons can travel a  longer distance along a wider jet before they escape via the
sides of the jet; again, they are less likely to hit the disk.

An interesting result of our work is the height distribution of the photons that
illuminate the disk. We have  found that the majority of photons that hit the
disk escape within a few gravitational radii, as expected in the lamp post model.
The exact range of heights depends on our choice of the variable $z_0$, which is the
distance from the center of the base of the jet to the black hole. However, there is a
significant contribution of photons that escape at larger radii
(Fig. \ref{height-dist-hitdisk}).  The large fraction of back-scattered   photons
at small heights $h$ is easily explained: In our model, we assume that photons
are injected at the sides of the jet at its base. At the bottom of the jet, the
optical depth is large (typically larger than 1). In addition,  the flow
velocity $v_{\parallel}$ is small (the acceleration zone extends up to $\sim
50\,r_g$). Hence, the photons do not experience a strong forward push. As a
result, many photons escape after one (or very few) scattering(s) and do not travel
long distances.   The cutoff at high heights is also easily explained: As the
height of the last scattering increases, the number of photons that hit the disk
decreases because the solid angle  that the disk subtends at this position decreases. 
Therefore, we expect a steep drop at high $h$, which is shown in
Fig.~\ref{height-dist-hitdisk}. Finally, the plateaus seen in
Fig.~\ref{height-dist-hitdisk} are due to the fact that the  optical depth in
these regions is of order unity and the photons are scattered with equal
probability there. 

The height distribution is fairly insensitive to changes in  optical depth
$\tau_{\parallel}$. Large values of the optical depth parallel to the jet axis
also imply large values of the optical depth perpendicular to the jet axis. The
two effects  compensate for each other, and photons travel  more or less the
same distance before they escape.  Indeed, the ratio
$\tau_{\parallel}/\tau_{\perp}$ as a function of $h$ is about the same 
irrespective of the values of the optical depths. In contrast, the height cutoff
decreases significantly as the jet width increases. This result can be explained
by the fact that photons escape from the sides more easily in a narrow jet. As
the width increases, more photons are able to travel along the jet before they
escape from the sides. But,  as mentioned above, the angle subtended by the disk
becomes smaller as the escaping height increases. Hence, photons that escape at
high heights  have a high chance of missing the disk.

The explanation of why the back-scattered spectrum is softer than the forward, 
direct one is also quite simple.  Consider soft photons entering a  scattering
region of an optical depth $\tau$ that is significantly larger than one. On
average, the photons penetrate the scattering region one mean free path 
(i.e., optical depth equal to one) . The photons that escape in the
forward direction encounter an optical depth $\tau -1$, while the photons that
are back-scattered encounter an optical depth of one.  The photons that escape
in the forward direction are scattered  more times than the back-scattered ones
and therefore gain more energy than the back-scattered ones.   This is because,
on average, the photons gain energy with every scattering. If there is bulk
flow  in the scattering region in the direction of the incoming photons, as is
the case of a jet, then this effect is stronger because the photons are pushed
in the forward direction by the bulk flow.

There is strong evidence that indicates that $R_F$ increases with luminosity
\citep{plant15,basak16,steiner16,walton17,wang-ji18,wang20}.  This result is
expected  in the truncated disk model because, as the luminosity increases -- that is to say, as
the system moves from the hard state to the intermediate and soft states -- the
inner disk radius decreases (the disk moves inward), and hence the area irradiated
in the disk increases. In Fig.~\ref{ref-frac-rad}, we naturally reproduce this
trend.

Finally, there is the crucial question of whether there is any observational
evidence for an extended and inhomogeneous lamp post. In this respect,
\citet{garcia19} showed that in order to achieve a good description of the
reflected spectrum of GX\,339--4, two sources of hard X-rays were needed: one 
located  a few gravitational radii from the black hole and the second lying at
$h\sim 600$ $r_g$. The second lamp post provided a better fit to the narrow
component of the Fe K emission. Similarly,  \citet{chakraborty20} obtained an
improvement in the model that fitted the observed spectrum by considering a
two-component corona at two different temperatures.  This difference in coronal
temperatures could be interpreted as originating from the fact that the two
components are located at different distances from the black hole. The
high-energy corona is much closer to the black hole and contributes to the broad
iron line through blurred reflection, while the low-energy corona is farther
away and contributes to the narrow core of the iron line complex. 
\citet{basak16} also found that, in the  lamp post model,  if the inner radius
of the accretion disk is fixed to the ISCO, then the height of the source of
hard X-rays is large (a few hundred $r_g$).

The report of a number of discrete lamp posts most likely reflects the fact that
the assumption of the lamp post model (an isotropic, stationary, point-like
source) is an idealized case of the real physical source, which is likely
extended, variable, and highly anisotropic \citep{dauser13}. Our model predicts
a continuum of values of $h$ at which the photons escape, and it appears to be a more
natural and physical representation.

\section{Conclusion}

Comptonization in the jet is inescapable.  The jet is fed from the hot inner
flow.  Thus, scattering in the hot inner flow, below the jet, will  result in
photons entering the jet and being scattered there as well. After all, the hot
inner flow and the base of the jet have no boundary.  Since in Comptonization it
is the last scatterings  that determine the outcome and not the first ones,
scattering in the  jet plays a fundamental role in shaping the 
radiation emitted by BHBs.
We have shown that a significant fraction of photons that are Comptonized in a
jet are back-scattered toward the accretion disk, despite the outward
relativistic bulk velocity  in the jet, and contribute to the illumination of
the disk. The jet appears to be an excellent candidate for the source of hard
X-rays in disk reflection models.

 

\bibliographystyle{aa}
\bibliography{../../../bhb.bib} 

\end{document}